\documentclass[prb,aps,twocolumn,showpacs,nobibnotes,epsf]{revtex4}
\usepackage{graphicx}% Include figure files
\usepackage{dcolumn}% Align table columns on decimal point
\usepackage{bm}% bold math
\usepackage{SIunits}

\begin{document}
\title{Evidence for Local Moment by Electron Spin Resonance Study on Polycrystalline LaFeAsO$_{1-x}$F$_x$ (x=0 and 0.13)}
\author{T. Wu, J. J. Ying, G. Wu, R. H. Liu, Y. He, H. Chen, X. F. Wang, Y. L. Xie, Y. J. Yan}
\author{X. H. Chen}
\altaffiliation{Corresponding author} \email{chenxh@ustc.edu.cn}
\affiliation{Hefei National Laboratory for Physical Science at
Microscale and Department of Physics, University of Science and
Technology of China, Hefei, Anhui 230026, People's Republic of
China\\}

\begin{abstract}
The temperature dependence of electron spin resonance (ESR) was
studied in the oxypnictide superconductors LaFeAsO$_{1-x}$F$_x$ (x =
0.0 and 0.13). In the samples, the ESR signal indicates that the g
factor and peak-to-peak linewidth strongly depend on temperature,
especially at low temperatures. It indicates a strong coupling
picture with existence of local moment. The dependence mentioned
above gradually attenuates, and tends to saturation around
room-temperature. This behavior could be ascribed to "bottleneck"
effect due to coupling of local moment and itinerant electron. In
addition, a Curie-Weiss like behavior is also observed in
temperature dependent integral intensity for the two samples. Our
results strongly support the existence of local moments in these
materials while its origin is still unclear. The results also
indicate strong magnetic frustration in this system, and magnetic
fluctuation mechanism for superconductivity is suggested.
\end{abstract}

\pacs{31.30.Gs,71.38.-k,75.30.-m}

\vskip 300 pt

\maketitle

The discovery of Fe-based high Tc superconductor provides a new
materials base to explore the mechanism of high-Tc superconductivity
besides high-Tc cuprates superconductor\cite{yoichi, chenxh, chen,
ren, rotter}. Similar to cuprates, Such pnictide superconductors are
also believed to have a quasi-2D conducting layer---Fe$_2$As$_2$
layer, which is separated by LnO (Ln = La, Sm etc.) or R (R = Ba, Sr
etc.) charge reservior. With doping electron or hole, the ground
state of FeAs compounds evolves from spin-density-wave state (SDW)
to superconducting state (SC). Electronic phase diagrams of FeAs
compounds were also found to be similar to the high-Tc
cuprates\cite{yoichi, dong, liu, Luetkens, zhao, Drew, chenhong},
where strong electron-electron coupling is believed to be the key to
understand high-Tc superconductivity. Therefore, one may naturally
ask whether it is still the case in iron-pnictides. As we know,
strong in-site Coulomb interaction can produce local moment in
cuprates. The existence of local moment is believed to be a strong
evidence for strong coupling. In FeAs parent compounds, the magnetic
moment of Fe$^{2+}$ ion is theoretically predicted to be 2.4 - 2.6
$\mu$$_B$\cite{Ma, cao, Ma2} while neutron results show a smaller
magnetic moment about 0.36-0.87$\mu$$_B$\cite{cruz, huang, Qiu,
zhao, zhao2}. Moreover, the static susceptibilities for parent
compounds decrease with decreasing temperature and shows a
linear-temperature behavior above SDW transition\cite{wang, gang,
Klingeler}. These results challenge the strong coupling picture. In
this paper, we $\textit{firstly}$ studied temperature dependence of
electron spin resonance (ESR) for LaFeAsO$_{1-x}$F$_x$ (x = 0.0 and
0.13). The benefit of ESR is that it can give us dynamic magnetic
information of local moment. An intrinsic resonance signal was
observed, and the g-factor and peak-to-peak linewidth show strong
temperature-dependence in low temperature region for the samples. It
indicates the existence of local moment, being consistent with
strong coupling picture. Further, "Bottleneck" effect due to
coupling between local moments and itinerant electrons was also
observed in high temperature region. In addition, a Curie-Weiss like
behavior in temperature dependent integral intensity was observed.
These results strongly support the strong coupling picture.

\begin{figure}[t]
\centering
\includegraphics[width=0.5\textwidth]{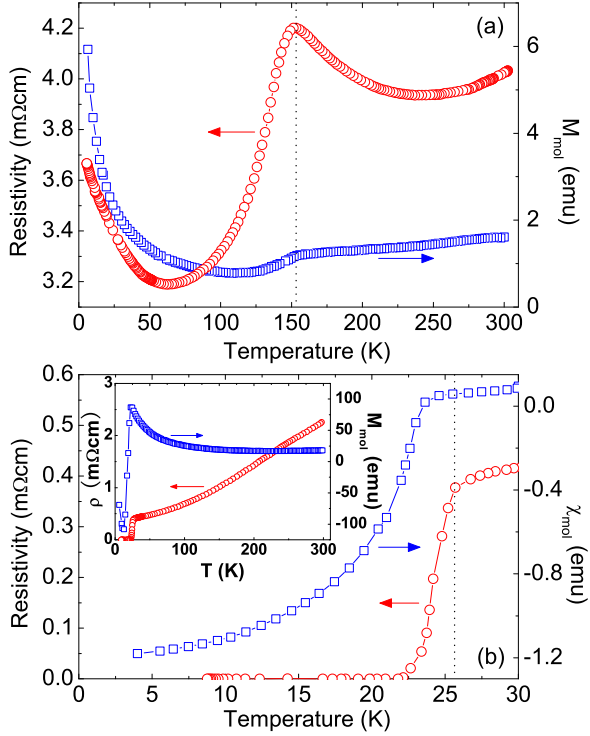}
\caption{(Color online). Temperature dependence of resistivity and
magnetization for LaFeAsO$_{1-x}$F$_x$ (a) x = 0.0 and (b) x =
0.13. Magnetization in (a) and inset of (b) are measured with H =
1 T and susceptibility in (b) is measured with H = 10 Oe.}
\label{fig1}
\end{figure}

Polycrystalline samples with nominal composition
$LaFeAsO_{1-x}F_{x}$ (x = 0, 0.13) were synthesized by
conventional solid state reaction using high purity LaAs, $LaF_3$,
Fe and $Fe_2O_3$ as starting materials. LaAs was obtained by
reacting La chips and As pieces at 600 $^oC$ for 3 hours and then
900 $^oC$ for 5 hours. The raw materials were thoroughly grounded
and pressed into pellets. The pellets were wrapped up by Ta foil
and sealed in an evacuated quartz tube. They were then annealed at
1160 $^oC$ for 40 hours. The sample preparation process except for
annealing was carried out in glove box in which high pure argon
atmosphere is filled. The XRD results show that the samples with
x=0 is single phase. A tiny but noticeable trace of impurity phase
LaOF was observed in x=0.13. The ESR measurements of the powder
samples were performed using a Bruker ER-200D-SRC spectrometer,
operating at X-band frequencies (9.47 GHz) and between 110 and 350
K. The resistance was measured by an AC resistance bridge (LR-700,
Linear Research). Magnetic susceptibility measurements were
performed with a superconducting quantum interference device
magnetometer in a magnetic field of 7 T. It should be addressed
that all results discussed as follow are well reproducible.

Fig.1 shows the temperature dependence of resistivity and
magnetization for LaFeAsO$_{1-x}$F$_x$ with x = 0 and 0.13. For
parent compound LaFeAsO, temperature dependence of resistivity shows
a peak around 155 K due to structural transition, and the
magnetization also shows a kink and at the same temperature as
reported previously\cite{yoichi}. For F-doped sample,
superconductivity with $T_c$ = 26 K was observed in resistivity and
magnetization as shown in Fig.1. These results show that the
electric and magnetic properties of samples used here is consistent
with previous works\cite{yoichi}, indicating a good starting for the
ESR study.

\begin{figure}[t]
%\captionstyle{flushleft} \onelinecaptionsfalse
\centering
\includegraphics[width=0.5 \textwidth]{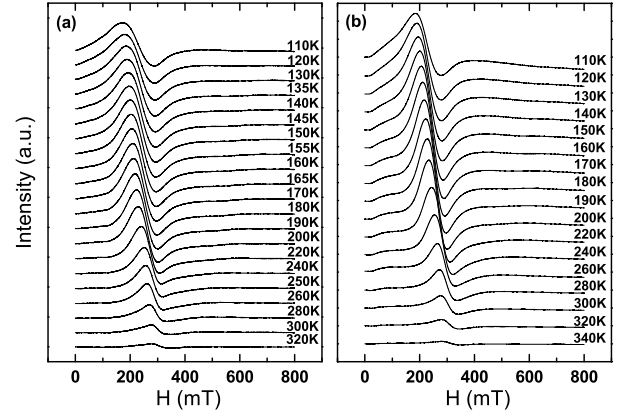}
\caption{ESR spectrum under different temperature for (a) x = 0.0
and (b) x = 0.13.} \label{fig2}
\end{figure}

Temperature dependences of ESR spectra for x = 0 and 0.13 samples in
the temperature range from 110 K to 350 K are shown in Fig.2. A tiny
background from secondary phase (eg. FeAs) in samples was subtracted
from the ESR spectra. A well-defined paramagnetic signal was
observed for the x = 0 and x = 0.13 samples. Lorentz formula was
used to fit the resonance signal very well. As shown in Fig.2, the
linewidth is broadened and the resonance field shifts to the lower
end with decreasing temperature. In traditional metal, no obvious
paramagnetic resonance signal is expected due to rapid spin-lattice
relaxation\cite{Taylor, Barnes}. The observed resonance here should
be considered to arise from local moment. In addition, we have
tested the possible effect from the impurities. Such paramagnetic
resonance signal was absent for all impurities at room temperature.
It proves that the observed signal is intrinsic for this system.

Fig.3 shows temperature dependence of g factor and peak-to-peak
linewidth ($\triangle$H$_{pp}$) for x = 0 and 0.13 samples.
$\triangle$H$_{pp}$ is defined as the width between the highest
point and the lowest point in temperature dependent ESR spectrum.
The resonance field (H$_c$) to calculate g factor is defined as the
magnetic field corresponding to the midpoint between the highest
point and the lowest point in the ESR spectrum. The g factor is
calculated by following formula: $g=\frac{h\nu}{\mu_BH_c}$. For the
parent compound, the g factor monotonously increases with decreasing
temperature below 300 K where a linear fitting works well. When
temperature exceeds 300 K, g-factor is saturated. In Fig.3(b), g
factor of F-doped sample shows a similar behavior. Compared to the
parent compound, a more clear trend of g-factor saturation is
presented above 260 K for the F-doped sample. Temperature dependent
$\triangle$H$_{pp}$ also shows a similar behavior for the two
samples. An upturn-behavior appears below 250 K for x = 0 sample and
200 K for x = 0.13 sample, respectively. Above the
upturn-temperature, the $\triangle$H$_{pp}$ is saturated with
increasing temperature in the two samples. The amplitude of
$\triangle$H$_{pp}$ for x = 0 is almost the same as that of x = 0.13
sample above upturn temperature, and becomes larger below upturn
temperature. It should be emphasized that the strong temperature
dependence of g factor and $\triangle$H$_{pp}$ cannot be accounted
for paramagnetic local moment because g$_{eff}$
(=g$_s$+$\triangle$g) is temperature-independent, while linewidth
1/$\tau$ (=a+bT) follows temperature-linear behavior for the system
with paramagnetic local moment. Strong temperature-dependent g
factor and $\triangle$H$_{pp}$ observed in ESR spectrum here has
been explained by magnetic fluctuation in the system with magnetic
phase transition\cite{Taylor, Barnes}. It suggests that magnetic
fluctuation from local moments exists in LaFeAsO$_{1-x}$F$_x$ (x =
0.0 and 0.13) system. Therefore, the question is naturally proposed:
How to understand the origin of local moment? One possible
explanation is that local spins from defects in FeAs layer lead to
paramagnetic resonance observed above. As shown in DC magnetization,
a Curie tail behavior is observed in the low temperature for parent
compound. By fitting the data with Curie-weiss formula, it is found
that the number of S = 1/2 local spin is about $\sim$ 0.02 per Fe
site and the Curie-weiss temperature is about $\sim$ 4 K. In this
situation, a very weak intensity of paramagnetic resonance is
expected at room temperature because the intensity is proportional
to DC magnetization from local spin (about $\sim$ 0.02 per Fe site).
Such expectation is in sharp contrast to the above observation. In
addition, g-factor is much larger than 2 as in free electron's case
as shown in Fig.3. Therefore, it indicates that the local spin from
defects can be ruled out, and the local moment should come from Fe
atom. But the magnetic state of Fe is still unclear. If local
moments exist in a metal, a so-called "bottleneck" effect takes
place in the transfer energy between the spin subsystem\cite{Taylor,
Barnes}. Figure 4 shows a cartoon picture for understanding
"bottleneck" effect. $\tau$$_{se}$ and $\tau$$_{es}$ are relaxation
times between local moments and itinerated electrons, respectively.
$\tau$$_{el}$ is spin-lattice relaxation time for itinerated
electrons. $\tau$$_{sl}$ is spin-lattice relaxation time for local
moments. Usually, the local spins are adiabatic for lattice and the
corresponding relaxation path is closed for local moments. When
$\tau$$_{se}$$>$$\tau$$_{el}$, the magnetic energy of local moment
can efficiently be passed on to lattice by itinerated electron, and
the effective relaxation of local moments is determined by
relaxation time $\tau$$_{se}$. When $\tau$$_{se}$$<$$\tau$$_{el}$,
magnetic energy of local moment, which is transferred to itinerated
electron, is quite likely to be returned back rather than passed on
to the lattice. Consequently, the relaxation process of the system
is dominated by slow relaxation $\tau$$_{el}$. The latter is called
"bottleneck" effect. This effect is successfully used to explain the
peculiar ESR features in $La_{1-x}Ca_xMnO_3$ and
La$_{2-x}$Sr$_x$CuO$_4$ system in which local moments and itinerated
electrons come from the same atoms\cite{Shenglaya, Kochelaev}. The
"bottleneck" effect is also expected in our system and can be used
to understand high-temperature behavior of g factor and
$\triangle$H$_{pp}$. The saturation of g factor and
$\triangle$H$_{pp}$ with increasing temperature indicates that
"bottleneck" effect gradually dominates with increasing temperature,
which is similar to the observed results in $La_{1-x}Ca_xMnO_3$ and
La$_{2-x}$Sr$_x$CuO$_4$\cite{Shenglaya, Kochelaev}. Since the
intensity of ESR signal decreases to the limit of the apparatus at
high temperatures, a more clear evidence of "bottleneck" effect is
lack in high temperature. But an increasing $\triangle$H$_{pp}$ and
almost invariable g factor are expected, which are observed in many
"bottleneck" system\cite{Shenglaya, Kochelaev}. At low temperatures,
the strong ferromagnetic fluctuation (g$\gg$2) frustrates the
relaxation between local moments and itinerated electrons and makes
corresponding relaxation slowing down. With decreasing temperature,
"bottleneck" effect is broken and a strong ferromagnetic fluctuation
between local moments enhances the g factor and $\triangle$H$_{pp}$.
These results show that there exists a ferromagnetic fluctuation
from local moments. However, an antiferrmagnetic order is
established below 135 K. It seems that dynamic magnetic properties
observed here are very different from static magnetic properties.

\begin{figure}[t]
%\captionstyle{flushleft} \onelinecaptionsfalse
\centering
\includegraphics[width=0.5\textwidth]{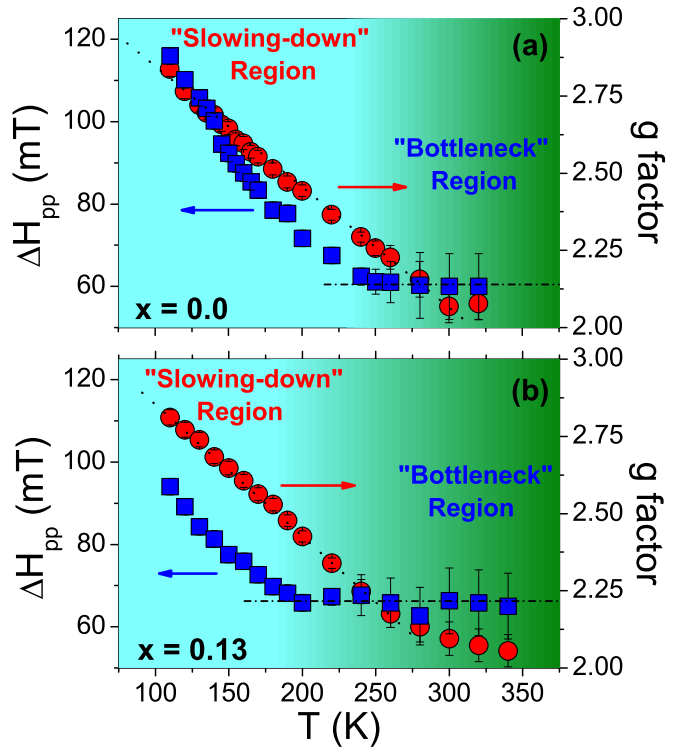}
\caption{(color online). Temperature dependence of g factor and
$\triangle$H$_{pp}$ for (a) x = 0.0 and (b) x = 0.13.}
\label{fig3}
\end{figure}

\begin{figure}[t]
%\captionstyle{flushleft} \onelinecaptionsfalse
\centering
\includegraphics[width=0.5\textwidth]{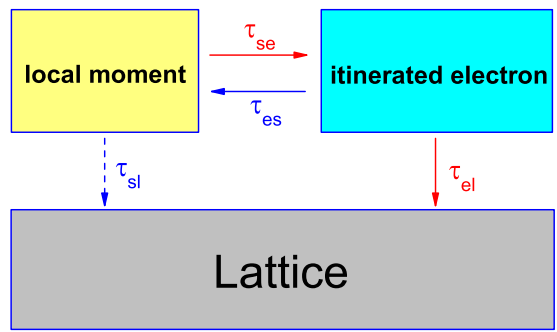}
\caption{(color online). Cartoon model for ESR ``bottleneck"
effect showing the various relaxation paths.} \label{fig4}
\end{figure}

\begin{figure}[t]
%\captionstyle{flushleft} \onelinecaptionsfalse
\centering
\includegraphics[width=0.5\textwidth]{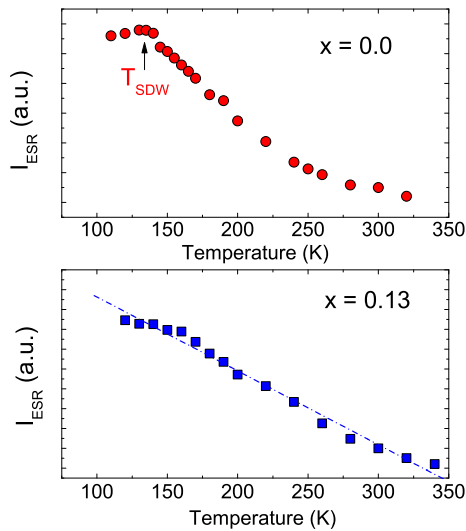}
\caption{(color online). Temperature dependence of integrated
intensity. Top panel: x = 0.0; Bottom panel: x = 0.13.}
\label{fig4}
\end{figure}

Fig.5 shows the temperature dependence of integral intensity of ESR
spectrum for both x = 0 and 0.13 samples. For x = 0 sample, a
Curie-weiss like behavior was observed for high temperature region
and a kink is observed around 135 K which corresponds to the
temperature of spin density wave transition determined by neutron
scattering. Below 135 K, the intensity of x=0 sample decreases with
decreasing temperature due to antiferromagnetic transition. For
F-doped superconducting sample, Curie-weiss like behavior is weaken
and a linear behavior is observed due to
I$_{ESR}$$\propto$$\chi$$_{ESR}$\cite{Taylor, Barnes}. Therefore,
this result is also consistent with local moments picture.

Recently, the origin of magnetic order in FeAs superconductors is
a very hot issue. Two distinct classes of theories have been
proposed: local moment antiferromagnetic ground state for strong
coupling\cite{cao, local2, local3, local4, local5, local6} and
itinerant ground state for weak coupling\cite{itinerant1,
itinerant2, itinerant3, itinerant4, itinerant5}. The local moment
magnetism approach stresses on-site correlations, and assumes that
the system is proximity to a Mott insulating state and the
resemblance to cuprates; while the latter approach emphasizes the
itinerated electron physics and the interplay between the
competing ferromagentic and antiferromagnetic fluctuation. The
observed results here may shed light on this debate. First, a
"local moment" effect is observed. "Bottleneck" effect,
Curie-weiss like behavior for $\chi_{ESR}$ and strong temperature
dependence of g factor and $\triangle$H$_{pp}$ strongly support
the existence of local moments. Secondly, ferromagnetic
fluctuation is observed among local moments. Our results seem to
favor strong coupling picture. But it is very strange that there
exists a ferromagnetic fluctuation among local moments above the
temperature of antiferromagnetic ordering. Kohama et al. observed
a large Wilson ratio in LaFeAsO$_{1-x}$F$_x$, indicating
ferromagnetic fluctuation in this system\cite{Kohama}. Recently,
Zhang et al. proposed a "performed SDW moment scenario" to explain
the temperature-linear behavior in DC magnetization\cite{zhang}.
It results from the existence of a wide fluctuation window in
which the local spin-density-wave correlation exists but the
global directional order has not yet been established. The
so-called local moment is defined as performed SDW moment in this
model. If we follow above idea, the observed ferromagnetic
coupling could be understand in the way that there exists a wide
ferromagnetic fluctuation window and it coexists and competes with
antiferromagnetic fluctuation. Therefore, strong magnetic
frustration maybe hidden in this system. Although the
$\triangle$H$_{pp}$ decreases with F-doping, a similar behavior of
g factor and $\triangle$H$_{pp}$ is observed in parent compound
and F-doped superconducting compound, and it indicates that strong
magnetic frustration is present in superconducting sample and
magnetic fluctuation may be very important to understand
superconductivity in this material. Similar result is also
obtained in DC magnetization for polycrystalline
LaFeAsO$_{1-x}$F$_x$\cite{Klingeler}.

In conclusion, we study temperature dependence of electron spin
resonance (ESR) for LaFeAsO$_{1-x}$F$_x$ (x = 0.0 and 0.13). Strong
temperature dependent g factor and $\triangle$H$_{pp}$ are observed
at low temperatures for the samples. Curie-weiss like behavior is
observed in the temperature dependent integral intensity. These
results strongly support the existence of local moments in these
materials, but its origin is still unclear (eg. "performed SDW
moment"). Strong magnetic frustration exists in both the parent
compound and superconducting sample. Magnetic fluctuation plays an
important role in mechanism for superconductivity.

This work is supported by the Nature Science Foundation of China
and by the Ministry of Science and Technology of China (973
project No: 2006CB601001) and by National Basic Research Program
of China (2006CB922005).

\end{document}